%% file: ResearchStat.tex
\documentclass{article}
\title{A Probabilistic Graph Model for Trust Opinion Estimation in Online Social Networks}
\author{Luke Liu and Qing Yang}
\date{}
\usepackage{nccfoots}
\usepackage{balance}
\usepackage[usenames,dvipsnames]{color}
\usepackage[tight,TABTOPCAP]{subfigure}
\usepackage{caption2}
\usepackage[cmex10]{amsmath}
\usepackage{multirow}
\usepackage{algorithmic}
\usepackage{float}
\usepackage{color, soul}
\usepackage{bm}
\usepackage[ruled,linesnumbered,vlined]{algorithm2e}
\usepackage{amsfonts}
\usepackage{graphicx}
\numberwithin{equation}{section}

\everymath{\displaystyle}

\begin{document}
%\fontsize{12}{15}
%\selectfont
\maketitle

{\sloppy
\input{content/abstract}
\input{content/introduction}

\input{content/relatedwork}

\input{content/background}

\input{content/model}

\input{content/Inference}

}

\bibliographystyle{unsrt}
\bibliography{./bib/ResearchStat}
\end{document}

%% file: content/abstract.tex
\begin{abstract}
Trust assessment plays a key role in many online applications, such as online money lending, product reviewing and active friending. Trust models usually employ a group of parameters to represent the trust relation between a trustor-trustee pair. These parameters are originated from the trustor’s bias and opinion on the trustee. Naturally, these parameters can be regarded as a vector. To address this problem, we propose a framework to accurately convert the single values to the parameters needed by 3VSL. The framework firstly employs a probabilistic graph model (PGM) to derive the trustor’s opinion and bias to his rating on the trustee.
\end{abstract}

%% file: content/introduction.tex
\section{Introduction}
%\label{sec:3vsl}
Trust assessment plays a key role in many online applications, such as online money lending, product reviewing and active friending. 
In previous work, we have proposed a model called 3VSL to assess the trust between unknown users though the trust social networks (TSNs) between people~\cite{6848107}. 
A TSN can be regarded as a graph, where the nodes are users and the edges are trust relations among the users. 
Each edge in a TSN is often associated with a single value or vector expressing the trust relation from the trustor to the trustee~\cite{5466696}. 
%

%
%The challenge is that the parameters of the DC distributions can hardly be estimated from most of the TSNs. 
%

%A TSN can be regarded a collection of trustor-trustee pairs. 
%
%Within each pair, there is a single value or vector representing the trust relation from the trustor to the trustee. 
%
Although 3VSL has been proved as an accurate model, applying it to most of the existing TSNs is still impractical.  
Particularly, 3VSL employs a group of parameters to represent the trust relation between a trustor-trustee pair. 
These parameters are originated from the trustor's bias and opinion on the trustee. 
Naturally, these parameters can be regarded as a vector. 
On the other hand, most of the TSNs employ an ordered rating, which is a single value, to represent the trust relation between a trustor-trustee pair.   
To apply 3VSL to these TSNs, previous work employs a heuristic manner to transform the single values to the parameters, which are not accurate.
As a result, the performance of 3VSL is seriously impacted.      

To address this problem, we propose a framework to accurately convert the single values to the parameters needed by 3VSL. 
The framework firstly employs a probabilistic graph model (PGM) to derive the trustor's opinion and bias to his rating on the trustee.
%
%While the ratings are the single values aforementioned in most of the TSNs,  the trustee's behavior and the trustor's bias are the parameters needed by 3VSL to model the trust between people.  
%
By using a Gibbs sampling algorithm upon the PGM, the parameters needed by 3VSL can be then estimated from the ratings existed in most of the TSNs.   
With the accurate parameters estimated by the framework, the performance of 3VSL is significantly boosted.   
Most importantly, with richer information provided by a group of values rather than a single value, we have opened a window for studying TSNs in a more informative way. 
In the following sections, we will introduce our PGM and Gibbs sampling algorithm for inference.  

%In this way, The model is able to estimate the parameters needed for 3VSL, from the single values provided in most of the TSNs. 
%
%We model the formation from a trustee's behavior to a trustor's rating on him as a PGM, where the parameters employed by 3VSL are latent variables  generating the rating.   
%

%% file: content/relatedwork.tex
\section{Related Work} 
\label{CH:RW}

How to model the trust between users in OSNs has attracted much attention in recent years.
Existing trust models can be categorized into four groups: topology based~\cite{wei2012sybildefender, danezis2009sybilinfer, yu2008sybilguard, 5313843}, PageRank based~\cite{Kamvar:2003:EAR:775152.775242, Gyongyi:2004:CWS:1316689.1316740}, probability based~\cite{despotovic2005probabilistic, teacy2012efficient,elsalamouny2010hmm,liu2012modeling,vogiatzis2010probabilistic}, and subjective logic based models~\cite{josang2001logic, 8737469, 7346861, 8057106, 6848107, 8613914, 8714017}.  
Along with the rapid development of the Internet and online services, trust has been used in many applications for either improving users' quality of experience (QoE) or preventing the disturbance of malicious users. Recently, trust was introduced in the concept of social cloud~\cite{8258338, 6682915, zhou2017will, 8622444, 8621918, assefi2016measuring, moyano2013re, assefi2015experimental, DiPietro201528}. Trust is also introduced in cyber-physical and edge computing systems, e.g., wireless sensor networks~\cite{6180964, 6142264, 8302985, 6040313, liu2019novel, Chen:2017:HEW:3136518.3047646, 7786108, 7236522, 7444932, chen2019understanding} and vehicular networks~\cite{SAT, 8567683, chen2019cooper, chen2019f}. Another important domain in which trust analysis is widely applied is Sybil defense and  spam detection~\cite{6547122, wei2012sybildefender, 8438939, 7346861, 5934998}. Details of the applications of trust can be seed in~\cite{8714017}
A detailed survey can be seen in~\cite{8714017}.

%% file: content/background.tex
\section{Background}
\label{sec:bkg}
In this section, we introduce the background knowledge of this paper.
\subsection{Behavior}
\label{sec:behv}
In 3VSL, a trustee's \textit{behavior} is modeled as a multinomial distribution upon interaction evidences, which contain $3$ possible categories: belief, distrust and neutral.  
Belief means the trustee will behave as is expected, distrust means the trustee will not behave as is expected, neutral means the trustee will either behave as is expected or not~\cite{Zou:2013:BPA:2505515.2507875}. 

Let's denote the parameters of trustor $j$'s behavior as $B_j = (b_j, d_j, n_j)$, where  $b_j$, $d_j$ and $n_j$ correspond to belief, distrust and neutral, respectively. 
Let's denote the interaction evidence collection $(\alpha_{ij}, \beta_{ij} ,\gamma_{ij})$ as the evidences trustor $i$ has observed from trustee $j$, where $\alpha_{ij}, \beta_{ij} ,\gamma_{ij}$ correspond to the numbers of events that trustee $j$ has behaved as expected, not as expected and unknown. 
Then, the counts of $\alpha_{ij}$, $\beta_{ij}$ and $\gamma_{ij}$ yield a multinomial distribution:
\begin{eqnarray}
{\alpha _{ij}},{\beta _{ij}},{\gamma _{ij}} \sim Mul({B_j},{\lambda _{ij}})
\label{behv}
\end{eqnarray}
where ${\lambda _{ij}} = {\alpha _{ij}} + {\beta _{ij}} + {\gamma _{ij}}$ denotes the total evidence number of $\o_{ij}$.
The pdf of $\alpha_{ij}, \beta_{ij} ,\gamma_{ij}$ can be expressed as
\begin{eqnarray}
&&P({\alpha _{ij}},{\beta _{ij}},{\gamma _{ij}}\left| {{b_j},{d_j},} \right.{n_j},{\lambda _{ij}}) \nonumber\\
&=& \frac{{\Gamma ({\lambda _{ij}} + 1)}}{{\Gamma ({\alpha _{ij}} + 1)\Gamma ({\beta _{ij}} + 1)\Gamma ({\gamma _{ij}} + 1)}}{({b_j})^{{\alpha _{ij}}}}{({d_j})^{{\beta _{ij}}}}{({n_j})^{{\gamma _{ij}}}}
\end{eqnarray}

%
%In Eq.~\ref{behv} and Eq.~\ref{multi}, $\alpha_j$, $\beta_j$ and $\gamma_j$ are the counts of events that trustee $j$ has behaves as expected, not as expected and unknown. 
%
%Parameter $|D|$ is the total number of the events.
%
%Parameters $b_j$, $d_j$ and $n_j$ equal to the expected proportions of events falling into the three statuses.   
%

\subsection{Opinion}
In 3VSL, an \textit{opinion} $\omega_{ij}$ is used to represent a trustee $j$'s trustworthiness from a certain trustor $i$'s view. 
It is written as:
\begin{equation}
{\omega_{ij}} = ({\alpha _{ij}},{\beta _{ij}},{\gamma _{ij}}){\left| a \right._i}
\label{opn}
\end{equation}
As shown in Eq.~\ref{opn}, an opinion $\omega_{ij}$ is composed of two parts. 

The left part, $({\alpha _{ij}},{\beta _{ij}},{\gamma _{ij}})$, denotes the counts of evidences $i$ has observed from his interactions with $j$. 
As shown previously, these evidences are observed by trustor $i$ from his interaction with trustee $j$.
Therefore, the counts of evidences from each category yield a multinomial distribution, as shown in Eq.~\ref{opn}.  
It is known that given evidences $({\alpha _{ij}},{\beta _{ij}},{\gamma _{ij}})$, the posterior of $B_{j}$ yields a Dirichlet distribution:
\begin{equation*}
{B^{* }_j} \sim Dir({\alpha _{ij}},{\beta _{ij}},{\gamma _{ij}})
\end{equation*}
It's pdf can be written as:
\begin{eqnarray}
P({B^{* }_j}\left| {{\alpha _{ij}},{\beta _{ij}},} \right.{\gamma _{ij}}) &=& P({b_{j}},{d_{j}},{n_{j}}\left| {{\alpha _{ij}},{\beta _{ij}},} \right.{\gamma _{ij}}) \hfill \nonumber \\
&=& \frac{{\Gamma ({\alpha _{ij}} + {\beta _{ij}} + {\gamma _{ij}})}}{{\Gamma ({\alpha _{ij}})\Gamma ({\beta _{ij}})\Gamma ({\gamma _{ij}})}}{({b_{j}})^{{\alpha _{ij}} - 1}}{({d_{j}})^{{\beta _{ij}} - 1}}{({n_{j}})^{{\gamma _{ij}} - 1}} \hfill  \nonumber \\
\label{Dir} 
\end{eqnarray}
Eq.~\ref{Dir} indicates that $B_{j}$ can be estimated from $i$'s opinion, which is originated from $i$'s interaction evidences with $j$. 
For simple, we denote $({\alpha _{ij}},{\beta _{ij}},{\gamma _{ij}})$ as $\o_{ij}$.
%The counts of the evidences for each category yield the multinomial distribution representing $j$'s behavior. 
%

The right part, $a_{i}$, is a float ranges from $0$ to $1$.
It represents $i$'s bias in evaluating $j$'s trustworthiness.   

Naturally, an opinion can be regarded as a vector.
3VSL employs an opinion to express the trust relation between a trustor-trustee pair. 
%3VSL employs the term $({\alpha _{ij}},{\beta _{ij}},{\gamma _{ij}})$ to estimate $B_j$ from $i$'s view. 
%
In this manner, the trust relation can be presented in a more informative manner, comparing to a single value.
%
%Then $j$'s trustworthiness in $i$'s view is computed by combining the estimated $B_j$ and $a_{i}$.
%
%Details of how to compute the trustworthiness will be introduced in the next session.  

\subsection{Expected Belief}
3VSL employs \textit{expected belief} to evaluate the total trustworthiness derived from an opinion. 
Taking the opinion $\omega_{ij}$ in Eq~\ref{opn} as an example, its expected belief $E({\omega_{ij}})$ is computed as:
\begin{eqnarray}
E({\o_{ij}}) = \frac{{{\alpha _{ij}} + {a_i}{\gamma _{ij}}}}{{{\alpha _{ij}} + {\beta _{ij}} + {\gamma _{ij}}}}
\label{expb}
\end{eqnarray}
In Eq.~\ref{expb}, the term $ \displaystyle \frac{{{\alpha _{ij}}}}{{{\alpha _{ij}} + {\beta _{ij}} + {\gamma _{ij}}}}$ accounts for the expectation of the belief part in an opinion.
The term $ \displaystyle \frac{{a_{i}{\gamma _{ij}}}}{{{\alpha _{ij}} + {\beta _{ij}} + {\gamma _{ij}}}}$ accounts for the expectation of belief in uncertainty.
Note that $a_{i}$ is used to determine how much belief can be accounted in uncertainty. 
In this paper, we assume that a trustworthiness rating $r_{ij}$ is derived from $i$'s expected belief on $j$.

\subsection{Ordered Logit Model}
\label{bkg:OLM}
An \textit{ordered logit model} is commonly used to transform a continuous variable to an ordered variable.    
Let the ordered levels be $1,2,...l,...L$. 
Let $x$ be a continuous independent variable and $y \in [1,L]$ be a variable dependent on $x$,
Then, the probability that the value of $y$ is greater than $l$ with respect to $x$ can be expressed as an ordered logit equation:
\begin{eqnarray}
P(y > l) &=& {\text{logit}}(\varphi x+ {\theta _l}) \hfill \nonumber \\
&=& \frac{{\exp (\varepsilon  x + {\theta_l})}}{{1 + \exp (\varepsilon  x + {\theta_l})}} \hfill \nonumber \\ 
\label{logit function}
\end{eqnarray}
%
%Eq.~\ref{logit function} can be explained as that which denotes the possible level $x$ belongs to.
%
Derived from Eq.~\ref{logit function}, the probability that $x$ belongs to level $l$ can be computed as:
\begin{eqnarray}
P(y = l) &=& \left\{ \begin{array}{l}
1 - P(y > l)\quad{\rm{ if }}\ l = 1\\
P(y > l - 1) - P(y > l)\quad {\rm{ if  }}\ 1 < l < L\\
P(y > l - 1)\quad{\rm{ if }}\ l = L
\end{array} \right.\nonumber \\
 &=& \left\{ \begin{array}{l}
1 - \displaystyle \frac{{\exp (\varepsilon {x} + {\theta_l})}}{{1 + \exp (\varepsilon {x} + {\theta_l})}}\quad{\rm{ if }}\ l = 1\\
\\
\displaystyle \frac{{\exp (\varepsilon {x} + {\theta_{l - 1}})}}{{1 + \exp (\varepsilon {x} + {\theta_{l-1}})}} - \frac{{\exp (\varepsilon {x} + {\theta_l})}}{{1 + \exp (\varepsilon {x} + {\theta_l})}}\quad{\rm{ if   }}\ 1 < l < L\\
\\
\displaystyle \frac{{\exp (\varepsilon {x} + {\theta_{l - 1}})}}{{1 + \exp (\varepsilon {x} + {\theta_{l - 1}})}}\quad{\rm{ if }}\ l = L
\end{array} \right. \nonumber\\
\label{logit_model_full}
\end{eqnarray}
where $\varepsilon$ and $\theta_l (1 \le l < L) $ are unknown parameters.
Note that $\left( {\sum\limits_{n \in [1,L]} {P(i = l)} } \right) = 1$.
Eq.~\ref{logit_model_full} is called ordered logit model. 
For simple, we rewrite it as the following form:
\begin{eqnarray}
y \sim {\rm{logit}}(\varepsilon x + \bm{\theta})
\label{logit_model}
\end{eqnarray} 
where $ \bm{\theta} = \theta_l (1 \le l < L)$

Given a collection of data couples $(x, y)$, the parameters $\varepsilon$ and $\theta_l (1 \le l < L) $ in Eq.~\ref{logit_model} can be estimated. 
As introduced before, an expected belief in 3VSL is a float ranges from $0$ to $1$, while a trustworthiness rating in most of the TSNs is an ordered variable. 
Therefore, our PGM employs ordered logit model to transform an expected belief to a trustworthiness rating, which will be detailed later.  
%realizes this transformation by estimating the values of ${\varphi}$ and ${\varepsilon _i}$, $\forall i \in 1,2,...$.
%
%Particularly, when there are $i = 4$ levels, we have:
%\begin{equation}
%\left\{ \begin{gathered}
 % P(i = 1) = 1 - \frac{{\exp (\varphi {x_i} + {\varepsilon _1})}}{{1 + \exp (\varphi {x_i} + {\varepsilon _1})}} \hfill \\
 % P(i = 2) = \frac{{\exp (\varphi {x_i} + {\varepsilon _1})}}{{1 + \exp (\varphi {x_i} + {\varepsilon _1})}} - \frac{{\exp (\varphi {x_i} + {\varepsilon _2})}}{{1 + \exp (\varphi {x_i} + {\varepsilon _2})}} \hfill \\
 % P(i = 3) = \frac{{\exp (\varphi {x_i} + {\varepsilon _2})}}{{1 + \exp (\varphi {x_i} + {\varepsilon _2})}} - \frac{{\exp (\varphi {x_i} + {\varepsilon _3})}}{{1 + \exp (\varphi {x_i} + {\varepsilon _3})}} \hfill \\
 % P(i = 4) = \frac{{\exp (\varphi {x_i} + {\varepsilon _3})}}{{1 + \exp (\varphi {x_i} + {\varepsilon _3})}} \hfill \\ 
%\end{gathered}  \right.
%\end{equation}    
\subsection{Gibbs Sampling}  
Gibbs sampling is a Markov chain Monte Carlo (MCMC) algorithm used to approximate a certain joint distribution, when the analytic form of the joint distribution cannot be solved out. 
The basic principle of Gibbs sampling is using large amount of samples extracted from various conditional distributions to approximate the joint distribution. 
Let the pdf of an unknown joint distribution be denoted as $P({x_1},{x_2},...{x_{i - 1}},{x_i},{x_{i + 1}}...{x_n})$, with respect to variables ${x_i},i \in [1,n]$.
Then, the Gibbs sampler algorithm can be written as:
\begin{algorithm}%[H]
\label{A1}
\caption{Gibbs Sampler}
\begin{algorithmic}[1]
\REQUIRE Randomly initialized variables $(x_1^{(0)},x_2^{(0)},...x_{i - 1}^{(0)},x_i^{(0)},x_{i + 1}^{(0)}....x_n^{(0)})$.
\ENSURE The distribution of data samples $(x_1^{(t)},x_2^{(t)},...x_{i - 1}^{(t)},x_i^{(t)},x_{i + 1}^{(t)}....x_n^{(t)})$, $t \in [0,T]$ converge when $T \to \infty$
\FORALL  {iteration number $l = 1,2,3...$}
\STATE $x_1^{(t)} \sim p(\left. {{x_1}} \right|x_2^{(t - 1)}...x_{i - 1}^{(t - 1)},x_i^{(t - 1)},x_{i + 1}^{(t - 1)}...x_n^{(t - 1)})$
\STATE $x_2^{(t)} \sim p(\left. {{x_2}} \right|x_1^{(t)}...x_{i - 1}^{(t - 1)},x_i^{(t - 1)},x_{i + 1}^{(t - 1)}...x_n^{(t - 1)})$
\STATE $\vdots$ 
\STATE $x_i^{(t)} \sim p(\left. {{x_i}} \right|x_1^{(t)}...x_{i - 1}^{(t)},x_{i + 1}^{(t - 1)}...x_n^{(t - 1)})$
\STATE $\vdots$ 
\STATE $x_n^{(t)} \sim p(\left. {{x_2}} \right|x_1^{(t)}...x_{i - 1}^{(t)},x_i^{(t)},x_{i + 1}^{(t)}...)$
\ENDFOR
\end{algorithmic}
\label{GS_arch}
\end{algorithm}
When the iteration number $t$ in Algorithm~\ref{GS_arch} is large enough, the distribution of conditional sampling couples $(x_1^{(t)},x_2^{(t)},...x_{i - 1}^{(t)},x_i^{(t)},x_{i + 1}^{(t)}....x_n^{(t)})$ for $t \in [0,T]$ will be the same as the data points sampled from  $P({x_1},{x_2},...{x_{i - 1}},{x_i},{x_{i + 1}}...{x_n})$. 
In this manner, the posterior of $P({x_1},{x_2},...{x_{i - 1}},{x_i},{x_{i + 1}}...{x_n})$ can be estimated upon the generated data points.

%% file: content/model.tex
\section{model}
\label{sec:3vsl}
In this section, we introduce the architecture of our PGM.

\subsection{Notations}
We first introduce the notations used in the rest of this paper. 

Let the collection storing all trustee's behaviors be denoted as a structured vector $B$, where $B_j \in B$ is an entry storing trustee $j$'s behavior parameters, i.e., $B_j = (b_j, d_j, n_j)$. 

Let the collection storing interaction evidence numbers between any trustor-trustee pair be denoted as a matrix $\lambda$, where $\lambda_{ij} \in n$ is an entry storing trustor $i$'s interaction evidence number with trustee $j$.

Let the collection storing interaction evidences between any trustor-trustee pair be denoted as a structured matrix $O$, where $\o_{ij} \in O$ is an entry storing trustor $i$'s interaction evidences with trustee $j$, i.e., $\o_{ij} = (\alpha_{ij}, \beta_{ij}, \gamma_{ij})$.

Let the collection storing all trustor's bias be denoted as a vector $a$, where $a_i \in a$ is an entry storing trustor $i$'s bias parameter, i.e., $a_i \in [0,1]$. 

Let an ordered logit model be denoted as $y \sim {\rm{logit}}(\varepsilon x + \bm{\theta} )$, where $\varepsilon$ and $\bm{\theta}$ are unknown parameters.

%Let the collection be denoted as $a$, where $a_i \in a$ is an entry storing trustor $i$'s bias parameter, i.e., $a_i \in [0,1]$. 
Let the collection storing trust ratings of all trustor-trustee pair be denoted as a matrix $R$, where $r_{ij} \in R$ is an entry storing trustor $i$'s trust rating on trustee $j$.

\subsection{Problem Statement}
We first formulate our problem, for which our PGM is about to addressing.
Let a rating network be a rating matrix $R$, where an entry $r_{ij}$  is an ordered rating indicating the trust strength from user $i$  to $j$. 
Let an opinion network be an opinion matrix  $O$, where an entry  $\omega_{ij}$ is an opinion from user $i$  to $j$. 
As is introduced before, the opinion $\omega_{ij}$ is a tuple of parameters determining a certain Dirichlet-Categorical distribution, which is employed by 3VSL to model the trust between a given trustor-trustee pair. 
Then, the problem is formulated as:

\textit{Given a rating matrix $R$, how to accurately estimate an opinion $\omega_{ij}$ for any $r_{ij} \in R$.}

\subsection{Model Architecture}
\label{model:arch}
We employ a hierarchical architecture to depict the transformation from an opinion $\o_{ij}$ between a trust pair to a corresponding rating $r_{ij}$.  
%
%In this session, we introduce the model upon which we transform the opinion between a trust pair to a corresponding rating. 
%

%
Firstly, the model takes trustee $j$'s interaction evidences under trustor $i$'s observation as a multinomial distribution, which can be expressed in Eq.~\ref{Mul}.
\begin{equation}
{\o_{ij}} \sim {\text{Mul}}({B_j},{\lambda_{ij}})
\label{Mul}
\end{equation}
As is introduced in Section~\ref{sec:behv}, $B_{j} = (b_j, d_j,n_j)$ are the parameters determining a trustor $j$'s behavior.
At the same time, $\lambda_{ij}$ is the parameter indicating the interaction evidences number between trustor $i$ and trustee $j$.
$\o_{ij} = (\alpha_{ij}, \beta_{ij} ,\gamma_{ij})$ indicate the counts of evidences belonging to belief, distrust and neutral.
Secondly, trustor $i$ forms his expected belief on $j$ upon the observed evidences and his personal bias, as shown in Eq.~\ref{model:expb}.
\begin{eqnarray}
E({\o _{ij}}) = \frac{{{\alpha _{ij}} + {a_i}{\gamma _{ij}}}}{{{\alpha _{ij}} + {\beta _{ij}} + {\gamma _{ij}}}}
\label{model:expb}
\end{eqnarray}
where $\alpha_{ij}$, $\beta_{ij}$ and $\gamma_{ij}$ are the evidences counts obtained previously, $a_i$ is the bias parameter introduced in Section~\ref{sec:expb}.

Finally, the expected belief becomes an ordered variable indicating the trust ratings presented in most of the TSNs.
We employ an ordered logit model to express this process, as shown in Eq.~\ref{logit_model}. 
\begin{eqnarray}
{r_{ij}} \sim {\rm{logit}}(\varepsilon E({\o _{ij}}) + \bm{\theta} )
\label{model:equation}
\end{eqnarray}

where $ \bm{{\theta}}$ indicates  $\theta_l (1 \le l < L) $ for different levels, as is presented in Section~\ref{bkg:OLM}.
%\begin{equation}
%{r_{ij}} \sim {\text{Probit}}(E({\o _{ij}}),\Theta ,\varepsilon )
%\end{equation}
%where $\Theta$ and $\varepsilon$ are parameters of the probit function.
%

\begin{figure}[!t]
\centering
\includegraphics[width=3in]{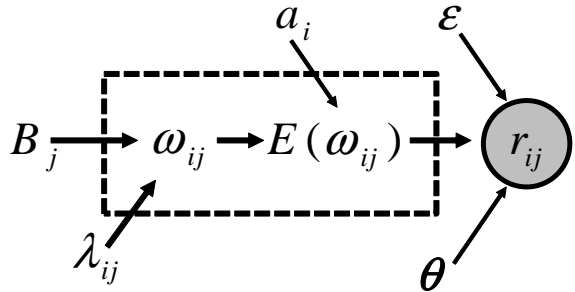}
\caption{Model of how trustor $i$ form his opinion and then rating on trustee $j$ from his behavior. The grey circle ($r_{ij}$) is rating, which is already observed. All of the others are parameters need to be estimated. Parameters in dashed rectangle are hidden variables. }
\label{fig:model}
\end{figure}

In summary, the complete model for generating $r_{ij} \in R$ is shown in Fig.~\ref{fig:model}.
Note that Fig.~\ref{fig:model} only depict the generation of a single entry in $R$. The whole rating matrix $R$ is formed as a joint results for all $r_{ij} \in R$.
For the whole model, only rating matrix $R$ is known while the rest of the parameters will be inferred. Details of the inference procedure will be introduced later.

%\begin{eqnarray}
%({\alpha _{ij}},{\beta _{ij}},{\gamma _{ij}}) \sim Mul({B_j},{n_{ij}})\nonumber\\
%{a_{ij}} \sim Uniform(0,1)\nonumber\\
%{e_{ij}} = {\alpha _{ij}} + {a_{ij}}{\gamma _{ij}}\nonumber\\
%{r_{ij}} \sim {\rm{logit}}(\varphi {e_{ij}} + \varepsilon )%\nonumber
%\label{eq:model}
%\end{eqnarray}

%% file: content/Inference.tex
\section{Inference}
\label{sec:Inf}

In this section, we introduce how to infer the unknown parameters in the model presented in Section~\ref{model:arch}. 
The approach we employed to conduct the inference is Gibbs sampling. 

\subsection{Model Formulation}

We first formulate our PGM introduced in Section~\ref{model:arch} upon a rating matrix $R$.
Let the pdf of the rating matrix $R$ with respect to parameters $B$,$\lambda$, $a$, $\bm{\theta}$ and $\epsilon$ be denoted as:
\begin{eqnarray}
P(R\left| {a,B,\lambda,\bm{\theta} ,\varepsilon } \right.)
\label{model_pdf}
\end{eqnarray} 
According to the Bayes' rule, the posterior estimation of parameters with respect to $R$ can be obtained from the following equation: 
\begin{eqnarray}
&&P(a,B,\lambda,\bm{\theta} ,\varepsilon \left| R \right.)\nonumber\\
 &=& \frac{{P(R,a,B,\lambda,\bm{\theta} ,\varepsilon )}}{{P(R)}}\nonumber\\
 &=& \frac{{P(R\left| {a,B,\lambda,\bm{\theta} ,\varepsilon } \right.)P(a,B,\lambda,\theta ,\varepsilon )}}{{P(R)}}
\label{posterior}
\end{eqnarray}
Eq.~\ref{posterior} shows that to estimate $B$,$\lambda$, $a$, $\bm{\theta}$ and $\epsilon$ from matrix $R$, we have to solve out $P(R\left| {a,B,\lambda,\bm{\theta} ,\varepsilon } \right.)$ and $P(a,B,\lambda,\bm{\theta} ,\varepsilon )$, respectively. 
%
%Since parameters $B$,$\lambda$, $a$, $\bm{\theta}$ and $\epsilon$ are independent with each other, for $P(a,B,\lambda,\bm{\theta} ,\varepsilon )$ we have:
%\begin{eqnarray}
%P(a,B,\lambda ,\bm{\theta} ,\varepsilon )\\
% &=& P(a)P(B)P(\lambda )P(\bm{\theta} )P(\varepsilon )\\
% &=& \left[ {\prod\limits_{i = 1}^m {P({a_i})} } \right]\left[ {\prod\limits_{j = 1}^n {P({B_j})} } \right]\left[ {\prod\limits_{i = 1}^m {\prod\limits_{j = 1}^n {P({\lambda _{ij}})} } } \right]\left[ {\prod\limits_{l = 1}^{L - 1} {P({\theta _l})} } \right]P(\varepsilon )
%\end{eqnarray}
%where $r_{ij} \in R$.
%
%As shown in Fig.~\ref{fig:model}, the factors contained by the model are: $B_j$,$n_{ij}$, $a_{ij}$, $\theta$ and $\epsilon$.
%
%The final result is rating matrix $R$.
%

We first try to solve out $P(R\left| {a,B,\lambda,\bm{\theta} ,\varepsilon } \right.)$.
According to Fig.~\ref{fig:model}, matrix $R$ is not directly determined by $B$ and $\lambda$. Instead, it is conditioned on a hidden variable matrix $O$, which is determined by parameters $B$ and $\lambda$.  
%
%Therefore, the probability that the rating matrix equals to $R$ and opinion matrix equals to $O$ is a joint distribution conditioned on factors $B$,$n$, $a$, $\theta$ and $\epsilon$:
%
%\begin{eqnarray}
%P(R,O\left| {a,B,n,\theta ,\varepsilon } \right.)
%\end{eqnarray}
%
%Since $O_{ij}$ is generated by multinomial distribution $Mul({B_j},n_{ij})$, the expression above can be rewritten as:
%
Therefore, Eq.~\ref{model_pdf} can be unfolded as: 
\begin{eqnarray}
\int {P(R\left| {O,a} \right.,\bm{\theta} ,\varepsilon )P(O\left| {B,\lambda} \right.)} dO
\label{model_pdf_uf}
\end{eqnarray}
We assume that each  $\o_{ij} \in O$ is independent with each other, and so does $r_{ij} \in R$.
As a result, the probability that matrix $R$ is observed equals to the product of probability that $r_{ij}$ is observed for all $r_{ij} \in R$. 
Similarly, the probability that matrix $O$ is observed equals to the product of probability that $\o_{ij}$ is observed for all $r_{ij} \in R$. 
Therefore, we have:
\begin{eqnarray}
&&P(R\left| {a,B,\lambda,\theta ,\varepsilon } \right.)\nonumber\\
&=& \int {P(R\left| {O,a} \right.,\theta ,\varepsilon )P(O\left| {B,\lambda} \right.)} dO\nonumber\\
&=& \int {\prod\limits_{i = 1}^m {\prod\limits_{j = 1}^n {P({r_{ij}}\left| {{\o_{ij}},{a_i}} \right.,\bm{\theta} ,\varepsilon )} } \prod\limits_{i = 1}^m {\prod\limits_{j = 1}^n {P({O_{ij}}\left| {{B_j},{\lambda_{ij}}} \right.)} } } dO
\label{product}
\end{eqnarray}

The term ${P({r_{ij}}\left| {{\o_{ij}},{a_i}} \right.,\bm{\theta} ,\varepsilon )}$ in Eq.~\ref{product} is the pdf of an ordered logit model, which can be denoted as:
\begin{eqnarray}
P({r_{ij}}\left| {{\o _{ij}},{a_i}} \right.,\theta ,\varepsilon ) = {\rm{logit}}(\varepsilon E({\omega _{ij}}) + \theta )
\label{Inf_logit}
\end{eqnarray}
where:
\begin{eqnarray}
E({\omega _{ij}}) = \frac{{{\alpha _{ij}} + {a_i}{\gamma _{ij}}}}{{{\alpha _{ij}} + {\beta _{ij}} + {\gamma _{ij}}}} \nonumber
\end{eqnarray}
Since our TSN has $4$ different levels, Eq.~\ref{Inf_logit} can be unfolded as:
%As is shown in Eq~\ref{logit_model_full}, 
\begin{eqnarray}
P({r_{ij}} = l) = \left\{ \begin{array}{l}
1 - \frac{{\exp \left[ {\varepsilon E({\omega _{ij}}) + {\theta _1}} \right]}}{{1 + \exp \left[ {\varepsilon E({\omega _{ij}}) + {\theta _1}} \right]}}\quad {\rm{ if  }} \ l = 1\nonumber\\
\frac{{\exp \left[ {\varepsilon E({\omega _{ij}}) + {\theta _1}} \right]}}{{1 + \exp \left[ {\varepsilon E({\omega _{ij}}) + {\theta _1}} \right]}} - \frac{{\exp \left[ {\varepsilon E({\omega _{ij}}) + {\theta _2}} \right]}}{{1 + \exp \left[ {\varepsilon E({\omega _{ij}}) + {\theta _2}} \right]}}\quad{\rm{ if  }} \ l = 2\nonumber\\
\frac{{\exp \left[ {\varepsilon E({\omega _{ij}}) + {\theta _2}} \right]}}{{1 + \exp \left[ {\varepsilon E({\omega _{ij}}) + {\theta _2}} \right]}} - \frac{{\exp \left[ {\varepsilon E({\omega _{ij}}) + {\theta _3}} \right]}}{{1 + \exp \left[ {\varepsilon E({\omega _{ij}}) + {\theta _3}} \right]}}\quad{\rm{if  }} \ l = 3\nonumber\\
\frac{{\exp \left[ {\varepsilon E({\omega _{ij}}) + {\theta _3}} \right]}}{{1 + \exp \left[ {\varepsilon E({\omega _{ij}}) + {\theta _3}} \right]}}\quad{\rm{if  }} \ l = 4\nonumber\\
\end{array} \right.
\end{eqnarray}

On the other hand, the term ${P({\o_{ij}}\left| {{B_j},{\lambda_{ij}}} \right.)}$ in Eq.~\ref{product} can be expressed as the pdf of a multinomial distribution:
\begin{eqnarray}
P({\o _{ij}}\left| {{B_j},{n_{ij}}} \right.) = \frac{{\Gamma ({\alpha _{ij}} + {\beta _{ij}} + {\gamma _{ij}} + 1)}}{{\Gamma ({\alpha _{ij}} + 1)\Gamma ({\beta _{ij}} + 1)\Gamma ({\gamma _{ij}} + 1)}}b_j^{{\alpha _{ij}}}d_j^{{\beta _{ij}}}n_j^{{\gamma _{ij}}}
\end{eqnarray}

Obviously, due to the complicated component within the integral in Eq.~\ref{product},  $P(R\left| {a,B,\lambda,\bm{\theta} ,\varepsilon } \right.)$ does not have an analytic form. 
As a result, we cannot directly obtain the maximum posterior estimation of Eq.~\ref{posterior} through derivation.
Therefore, we use Gibbs sampling algorithm to approach the approximate distribution of ${P(R,a,B,\lambda,\bm{\theta} ,\varepsilon )}$, with respect to the hidden variable $O$.  

\subsection{Conditional Probability}
As is shown in Algorithm~\ref{GS_arch}, to setup a Gibbs sampling algorithm, we need to know the pdf of each parameter conditioned on the rest parameters.    
By accounting for the hidden variable $O$, the conditional pdfs we are interested in are:  
\begin{eqnarray}
P(O\left| {R,a,B,\lambda, \bm{\theta} ,\varepsilon } \right.)\nonumber\\
P(a\left| {R,O,B,\lambda,\bm{\theta} ,\varepsilon } \right.)\nonumber\\
P(B\left| {R,O,a,\lambda,\bm{\theta} ,\varepsilon } \right.)\nonumber\\
P(\lambda\left| {R,O,a,B,\bm{\theta} ,\varepsilon } \right.)\nonumber\\
P(\bm{\theta} \left| {R,O,a,B,\lambda,\varepsilon } \right.)\nonumber\\
P(\varepsilon \left| {R,O,a,B,\lambda, \bm{\theta} } \right. )\nonumber
\end{eqnarray}
%$P(O\left| {R,a,B,n,\theta ,\varepsilon } \right.)$, $P(a\left| {R,O,B,n,\theta ,\varepsilon } \right.)$, $P(B\left| {R,O,a,n,\theta ,\varepsilon } \right.)$, $P(n\left| {R,O,a,B,\theta ,\varepsilon } \right.)$, $P(\theta \left| {R,O,a,B,n,\varepsilon } \right.)$ and $P(\varepsilon \left| {R,O,a,B,n,} \right.\theta )$.
%
We will solve out each of them in the rest of this section.  
%

%\subsubsection{$P(O\left| {R,a,B,n,\theta ,\varepsilon } \right.)$}
\subsubsection{$P(O\left| {R,a,B,\lambda,\bm{\theta} ,\varepsilon } \right.)$}
We first solve $P(O\left| {R,a,B,\lambda,\bm{\theta} ,\varepsilon } \right.)$. 
Since $O$ is a structured matrix where an entry is denoted as $\o_{ij} = (\alpha_{ij}, \beta_{ij}, \gamma_{ij})$, we need to further solve:
\begin{eqnarray}
P({\o_{ij}}\left| {R,O/{\o_{ij}},a,B,\lambda,\bm{\theta} ,\varepsilon } \right.)
\label{O_ij}
\end{eqnarray}
for each $\o_{ij} \in O$, where ${O/{\o_{ij}}}$ denotes all the other entries in $O$ except $\o_{ij}$.
According to Bayes' rule, Eq.~\ref{O_ij} can also be expressed as:
\begin{eqnarray}
&&P({\o_{ij}}\left| {O/{\o_{ij}},a,B,\lambda, \bm{\theta}, R} \right.)\nonumber\\
&=& \frac{\displaystyle {P(R,O,a,B,\lambda,\bm{\theta} ,\varepsilon )}}{{P(R,O/{\o_{ij}},a,B,\lambda,\bm{\theta} ,\varepsilon )}}\nonumber\\
%\label{O_ij_joint} 
& =& \frac{\displaystyle {P(R\left| {O,a,\bm{\theta} ,\varepsilon } \right.)P(O\left| {B,\lambda} \right.)}}{{P(R\left| {O/{\o_{ij}},a,\theta ,\varepsilon } \right.)P(O/{\o_{ij}}\left| {B,\lambda} \right.)}} \nonumber \\
& =& \frac{\displaystyle {P(R\left| {O,a,\bm{\theta} ,\varepsilon } \right.)}}{{P(R\left| {O/{\o_{ij}},a,\bm{\theta} ,\varepsilon } \right.)}}\frac{{P(O\left| {B,\lambda} \right.)}}{{P(O/{\o_{ij}}\left| {B,\lambda} \right.)}}
\label{O_ij_condition}
\end{eqnarray}
%
%Note that the term $P(R,O/{\o_{ij}},a,B,n,\theta ,\varepsilon )$ in Eq.~\ref{O_ij_joint} is another writing form of $P({\o_{ij}},R,O/{O_{ij}},a,B,n)$. 
%

For the term $\displaystyle \frac{{P(O\left| {B,\lambda} \right.)}}{{P(O/{\o_{ij}}\left| {B,\lambda} \right.)}}$ in Eq.~\ref{O_ij_condition}, we simplify it as followings:
\begin{eqnarray}
&&\frac{{P(O\left| {B,\lambda } \right.)}}{{P(O/{\o _{ij}}\left| {B,\lambda } \right.)}}\nonumber\\
 &=& \frac{{\prod\nolimits_{n = 1}^N {\prod\nolimits_{m = 1}^M {P({\o _{mn}}\left| {{B_n},{\lambda _{mn}}} \right.)} } }}{{\prod\nolimits_{n = 1}^N {\prod\nolimits_{m = 1}^{M/i} {P({\o _{mn}}\left| {{B_n},{\lambda _{mn}}} \right.)} } }}\nonumber\\
 &=& P({\o _{ij}}\left| {{B_j},{\lambda _{ij}}} \right.)\nonumber\\
 &=& \frac{{\Gamma ({\alpha _{ij}} + {\beta _{ij}} + {\gamma _{ij}} + 1)}}{{\Gamma ({\alpha _{ij}} + 1)\Gamma ({\beta _{ij}} + 1)\Gamma ({\gamma _{ij}} + 1)}}b_j^{{\alpha _{ij}}}d_j^{{\beta _{ij}}}n_j^{{\gamma _{ij}}}
\label{O_ij_B}
\end{eqnarray}
Eq.~\ref{O_ij_B} indicates that any $\o_{ij} \in O$ is determined by a multinomial distribution with paramters $B_j$ and $\lambda_{ij}$. 

For the term $\displaystyle \frac{{P(R\left| {O,a,\bm{\theta} ,\varepsilon } \right.)}}{{P(R\left| {O/{\o_{ij}},a,\bm{\theta} ,\varepsilon } \right.)}}$ in Eq.~\ref{O_ij_condition}, we unfold it as followings:
\begin{eqnarray}
&&\frac{{P(R\left| {O,a} \right.,\bm{\theta} ,\varepsilon )}}{{P(R\left| {O/{\o _{ij}},a} \right.,\bm{\theta} ,\varepsilon )}}\nonumber\\
 &=& \frac{{\prod\nolimits_{n = 1}^N {\prod\nolimits_{m = 1}^M {{\rm{logit}}\left[ {\varepsilon E({\omega _{mn}}) + \bm{\theta} } \right]} } }}{{\left( {\sum\nolimits_{{\o _{ij}}} {{\rm{logit}}\left[ {\varepsilon E({\omega _{ij}}) + \bm{\theta} } \right]} } \right)\prod\nolimits_{n = 1}^N {\prod\nolimits_{m = 1}^{M/i} {{\rm{logit}}\left[ {\varepsilon E({\omega _{mn}}) + \theta } \right]} } }}\nonumber\\
& =& \frac{{{\rm{logit}}\left[ {\varepsilon E({\omega _{ij}}) +\bm{\theta} } \right]}}{{\sum\nolimits_{{\o _{ij}}} {{\rm{logit}}\left[ {\varepsilon E({\omega _{ij}}) + \bm{\theta} } \right]} }}
\label{O_ij_R} 
\end{eqnarray}
where $\sum\nolimits_{{\o _{ij}}} $ denotes the summation for all possible values of $\o_{ij}$, ${{\rm{logit}}\left[ {\varepsilon E({\omega _{ij}}) + \bm{\theta} } \right]}$ is unfolded as shown in Eq.~\ref{Inf_logit}.
Combine Eq.~\ref{O_ij_R} and~\ref{O_ij_B} together, the analytic form of $P(O\left| {R,a,B,n,\theta ,\varepsilon } \right.)$ is eventually expressed as:
\begin{eqnarray}
&&P(O\left| {R,a,B,\lambda,\bm{\theta} ,\varepsilon } \right.) \nonumber\\
&=& \frac{{{\rm{logit}}\left[ {\varepsilon E({\omega _{ij}}) +\bm{\theta} } \right]}}{{\sum\nolimits_{{\o _{ij}}} {{\rm{logit}}\left[ {\varepsilon E({\omega _{ij}}) + \bm{\theta} } \right]} }}\frac{{\Gamma ({\alpha _{ij}} + {\beta _{ij}} + {\gamma _{ij}} + 1)}}{{\Gamma ({\alpha _{ij}} + 1)\Gamma ({\beta _{ij}} + 1)\Gamma ({\gamma _{ij}} + 1)}}b_j^{{\alpha _{ij}}}d_j^{{\beta _{ij}}}n_j^{{\gamma _{ij}}}\nonumber\\
\label{O_ij_final}
\end{eqnarray}

\subsubsection{$P(B\left| {R,O,a,\lambda,\bm{\theta} ,\varepsilon } \right.)$}

For $P(B\left| {R,O,a,\lambda,\bm{\theta} ,\varepsilon } \right.)$, since $B$ is a structured vector where an entry is denoted as $B_{j}$, we need to further solve out
\begin{eqnarray}
P({B_j}\left| {R,O,B/{B_j},a,\lambda,\bm{\theta} ,\varepsilon } \right.)
\label{B_j}
\end{eqnarray}
for each $B_{j} \in B$, where ${B/{B_{j}}}$ denotes all the other entries in $B$ except $B_{j}$.

According to Bayes' rule, Eq.~\ref{B_j} can be rewritten as:
\begin{eqnarray}
&&P(R,O,a,B,n,\theta ,\varepsilon )\nonumber\\
& =& \prod\nolimits_{n = 1}^N {\prod\nolimits_{m = 1}^M {P({r_{mn}}\left| {{\o _{mn}},{a_m},\theta ,\varepsilon } \right.)P({\o _{mn}}\left| {{B_n},{\lambda _{mn}}} \right.)} } 
\label{B_ij_condition}
\end{eqnarray}
The numerator in Eq.~\ref{B_ij_condition} can be unfolded as:
\begin{eqnarray}
&&P(R,O,a,B,\lambda,\bm{\theta} ,\varepsilon )\\
 &=& \prod\nolimits_{n = 1}^N {\prod\nolimits_{m = 1}^M {P({r_{ij}}\left| {{\o _{ij}},{a_i},\bm{\theta} ,\varepsilon } \right.)P({\o _{ij}}\left| {{B_j},{\lambda _{ij}}} \right.)} }
 \label{B_num}
\end{eqnarray}
while the denominator in Eq.~\ref{B_ij_condition} can be unfolded as:
\begin{eqnarray}
&&P(R,B/{B_j},a,B,\lambda,\bm{\theta} ,\varepsilon )\nonumber\\
 &=& \left[ {\prod\nolimits_{n = 1}^{N/j} {\prod\nolimits_{m = 1}^M {P({r_{mn}}\left| {{\o _{mn}},{a_m},\bm{\theta} ,\varepsilon } \right.)P({\o _{mn}}\left| {{B_n},{\lambda _{mn}}} \right.)} } } \right] \times \nonumber\\
&&\int {\prod\nolimits_{m = 1}^M {P({r_{mj}}\left| {{\o _{mj}},{a_m},\bm{\theta} ,\varepsilon } \right.)P({\o _{mj}}\left| {{B_j},{n_{mj}}} \right.)} d{B_j}} 
\label{B_den}
\end{eqnarray}

By substituting the numerator and denominator in Eq.~\ref{B_ij_condition} with Eq.~\ref{B_num} and~\ref{B_den} and eliminate the redundant terms, we get:
\begin{eqnarray}
&&\frac{ {P(R,O,a,B,\lambda ,\bm{\theta} ,\varepsilon )}}{{P(R,O,a,B/{B_j},\lambda ,\bm{\theta} ,\varepsilon )}}\nonumber\\
 &=& \frac{{\displaystyle \prod\nolimits_{m = 1}^M {P({r_{mj}}\left| {{\o _{mj}},{a_m},\bm{\theta} ,\varepsilon } \right.)P({\o _{mj}}\left| {{B_j},{\lambda _{mj}}} \right.)} }}{{\displaystyle \int {\prod\nolimits_{m = 1}^M {P({r_{mj}}\left| {{\o _{mj}},{a_m},\bm{\theta} ,\varepsilon } \right.)P({\o _{mj}}\left| {{B_j},{\lambda _{mj}}} \right.)} } d{B_j}}}\nonumber\\
 &=& \frac{{\displaystyle\prod\nolimits_{m = 1}^M {P({r_{mj}}\left| {{\o _{mj}},{a_m},\bm{\theta} ,\varepsilon } \right.)} \prod\nolimits_{m = 1}^M {P({\o _{mj}}\left| {{B_j},{\lambda _{mj}}} \right.)} }}{{\displaystyle \prod\nolimits_{m = 1}^M {P({r_{mj}}\left| {{\o _{mj}},{a_m},\bm{\theta} ,\varepsilon } \right.)} \displaystyle \int {\prod\nolimits_{m = 1}^M {P({\o _{mj}}\left| {{B_j},{\lambda _{mj}}} \right.)} d{B_j}} }}\nonumber\\
 &=& \frac{{\displaystyle\prod\nolimits_{m = 1}^M {P({\o _{mj}}\left| {{B_j},{\lambda _{mj}}} \right.)} }}{{ \displaystyle \int {\prod\nolimits_{m = 1}^M {P({\o _{mj}}\left| {{B_j},{\lambda _{mj}}} \right.)} d{B_j}} }}
\label{B_middle}
\end{eqnarray}

%By unfolding Eq.~\ref{B_ij_condition}, then we have:
%
%\begin{eqnarray}
%&&\frac{{P(R,O,a,B,n,\theta ,\varepsilon )}}{{P(R,B/{B_j},a,B,n,\theta ,\varepsilon )}}\nonumber\\
 %&=& \frac{{\prod\limits_{n = 1}^N {\prod\limits_{m = 1}^M {P({R_{ij}}\left| {{O_{ij}},{a_i},\theta ,\varepsilon } \right.)P({O_{ij}}\left| {{B_j},{n_{ij}}} \right.)} } }}{{\prod\limits_{j = 1}^{n/j} {\prod\limits_{i = 1}^m {P({R_{ij}}\left| {{O_{ij}},{a_i}} \right.,\theta ,\varepsilon )P({O_{ij}}\left| {{B_j},{n_{ij}}} \right.)} } \displaystyle \int {\prod\limits_{i = 1}^m {P({R_{ij}}\left| {{O_{ij}},{a_i},\theta ,\varepsilon } \right.)P({O_{ij}}\left| {{B_j},{n_{ij}}} \right.)} d{B_j}} }}\nonumber\\
 %&=& \frac{{\prod\limits_{i = 1}^m {P({R_{ij}}\left| {{O_{ij}},{a_i},\theta ,\varepsilon } \right.)} P({O_{ij}}\left| {{B_j},{n_{ij}}} \right.)}}{{\displaystyle \int {\prod\limits_{i = 1}^m {P({R_{ij}}\left| {{O_{ij}},{a_i},\theta ,\varepsilon } \right.)} P({O_{ij}}\left| {{B_j},{n_{ij}}} \right.)} d{B_j}}}\nonumber\\
 %&=& \frac{{\prod\limits_{i = 1}^m {P({R_{ij}}\left| {{O_{ij}},{a_i},\theta ,\varepsilon } \right.)} \prod\limits_{i = 1}^m {P({O_{ij}}\left| {{B_j},{n_{ij}}} \right.)} }}{{\prod\limits_{i = 1}^m {P({R_{ij}}\left| {{O_{ij}},{a_i}} \right.,\theta ,\varepsilon )} \displaystyle \int {\prod\limits_{i = 1}^m {P({O_{ij}}\left| {{B_j},{n_{ij}}} \right.)} d{B_j}} }}\nonumber\\
 %&=& \frac{{\prod\limits_{i = 1}^m {P({O_{ij}}\left| {{B_j},{n_{ij}}} \right.)} }}{{\displaystyle \int {\prod\limits_{i = 1}^m {P({O_{ij}}\left| {{B_j},{n_{ij}}} \right.)} d{B_j}} }}
%\label{B_} 
%\end{eqnarray}
%

Since ${P({\o_{ij}}\left| {{B_j},{\lambda_{ij}}} \right.)}$ is the pdf a multinomial distribution, the numerator of Eq.~\ref{B_middle} can be further unfolded as:
\begin{eqnarray}
&&\prod\nolimits_{m = 1}^M {P({\o _{mj}}\left| {{B_j},{\lambda _{mj}}} \right.)} \nonumber\\
 &=& \prod\nolimits_{m = 1}^M {\displaystyle \frac{{\Gamma ({\alpha _{mj}} + {\beta _{mj}} + {\gamma _{mj}} + 1)}}{{\Gamma ({\alpha _{mj}} + 1)\Gamma ({\beta _{mj}} + 1)\Gamma ({\gamma _{mj}} + 1)}}b_j^{{\alpha _{mj}}}d_j^{{\beta _{mj}}}n_j^{{\gamma _{mj}}}} 
 \label{B_middle_num}
\end{eqnarray}
while the denominator of Eq.~\ref{B_middle} can be further unfold as:
\begin{eqnarray}
&&\int {\prod\nolimits_{m = 1}^M {P({\o _{mj}}\left| {{B_j},{\lambda _{mj}}} \right.)} d{B_j}} \nonumber\\
 &=& \int {\prod\nolimits_{m = 1}^M \displaystyle {\frac{{\Gamma ({\alpha _{mj}} + {\beta _{mj}} + {\gamma _{mj}} + 1)}}{{\Gamma ({\alpha _{mj}} + 1)\Gamma ({\beta _{mj}} + 1)\Gamma ({\gamma _{mj}} + 1)}}b_j^{{\alpha _{mj}}}d_j^{{\beta _{mj}}}n_j^{{\gamma _{mj}}}} d{B_j}} \nonumber\\
 &=& \left[ {\prod\nolimits_{m = 1}^M \displaystyle {\frac{{\Gamma ({\alpha _{mj}} + {\beta _{mj}} + {\gamma _{mj}} + 1)}}{{\Gamma ({\alpha _{mj}} + 1)\Gamma ({\beta _{mj}} + 1)\Gamma ({\gamma _{mj}} + 1)}}} } \right] \times \nonumber\\
&&\int {\left( {b_j^{\sum\nolimits_{m = 1}^M {{\alpha _{mj}}} }d_j^{\sum\nolimits_{m = 1}^M {{\beta _{mj}}} }n_j^{\sum\nolimits_{m = 1}^M {{\gamma _{mj}}} }} \right)d{B_j}} 
 \label{B_middle_den}
\end{eqnarray}
By substituting the numerator and denominator of Eq.~\ref{B_middle} with Eq.~\ref{B_middle_num} and Eq.~\ref{B_middle_den} and eliminating the redundant terms, we have:
\begin{eqnarray}
&&\frac{{\displaystyle \prod\nolimits_{m = 1}^M {P({\o _{mj}}\left| {{B_j},{\lambda _{mj}}} \right.)} }}{{\displaystyle\int {\prod\nolimits_{m = 1}^M {P({\o _{mj}}\left| {{B_j},{\lambda _{mj}}} \right.)} d{B_j}} }}\nonumber\\
&=& \frac{\displaystyle{\left( {b_j^{\sum\nolimits_{m = 1}^M {{\alpha _{mj}}} }d_j^{\sum\nolimits_{m = 1}^M {{\beta _{mj}}} }n_j^{\sum\nolimits_{m = 1}^M {{\gamma _{mj}}} }} \right)}}{{\displaystyle\int {\left( {b_j^{\sum\nolimits_{m = 1}^M {{\alpha _{mj}}} }d_j^{\sum\nolimits_{m = 1}^M {{\beta _{mj}}} }n_j^{\sum\nolimits_{m = 1}^M {{\gamma _{mj}}} }} \right)d{B_j}} }}\nonumber \\
 &=& \frac{\displaystyle {\left( {b_j^{\sum\nolimits_{m = 1}^M {{\alpha _{mj}}} }d_j^{\sum\nolimits_{m = 1}^M {{\beta _{mj}}} }n_j^{\sum\nolimits_{m = 1}^M {{\gamma _{mj}}} }} \right)}}{{\displaystyle \frac{{\Gamma (1 + \sum\nolimits_{i = 1}^M {{\alpha _{mj}}} )\Gamma (1 + \sum\nolimits_{i = 1}^M {{\beta _{mj}}} )\Gamma (1 + \sum\nolimits_{i = 1}^M {{\gamma _{mj}}} )}}{{\Gamma \left[ {(1 + \sum\nolimits_{i = 1}^M {{\alpha _{mj}}} ) + (1 + \sum\nolimits_{i = 1}^M {{\beta _{mj}}} ) + (1 + \sum\nolimits_{i = 1}^M {{\gamma _{mj}}} )} \right]}}}}\nonumber\\
 &=& \displaystyle \left\{ {\frac{{\Gamma \left[ {3 + \sum\nolimits_{i = 1}^M {{\alpha _{mj}}}  + \sum\nolimits_{i = 1}^M {{\beta _{mj}}}  + \sum\nolimits_{i = 1}^M {{\gamma _{mj}}} } \right]}}{{\Gamma (1 + \sum\nolimits_{i = 1}^M {{\alpha _{mj}}} )\Gamma (1 + \sum\nolimits_{i = 1}^M {{\beta _{mj}}} )\Gamma (1 + \sum\nolimits_{i = 1}^M {{\gamma _{mj}}} )}}} \right\} \times \\
&&\left( {b_j^{\sum\nolimits_{m = 1}^M {{\alpha _{mj}}} }d_j^{\sum\nolimits_{m = 1}^M {{\beta _{mj}}} }n_j^{\sum\nolimits_{m = 1}^M {{\gamma _{mj}}} }} \right)\nonumber
\label{B_final}
\end{eqnarray}
Eq.~\ref{B_final} is the analytic form of $P({B_j}\left| {R,O,B/{B_j},a,\lambda,\bm{\theta} ,\varepsilon } \right.)$.

\subsubsection{$P(\lambda\left| {R,O,a,B,\bm{\theta} ,\varepsilon } \right.)$}
For $P(\lambda\left| {R,O,a,B,\bm{\theta} ,\varepsilon } \right.)$, since $\lambda$ is a matrix where an entry is denoted as $\lambda_{ij}$, we need to further solve out:
\begin{eqnarray}
P({\lambda_{ij}}\left| {R,O,B/{B_j},a,n, \bm{\theta}, \varepsilon } \right.)
\label{n_ij}
\end{eqnarray}
for each $\lambda_{ij} \in \lambda$. 
Note that ${\lambda / \lambda_{ij} }$ denotes all the other entries in $\lambda$ except $\lambda_{ij}$.
According to Bayes' rule, Eq.~\ref{n_ij} can be written as:
\begin{equation}
\begin{array}{l}
P({\lambda_{ij}}\left| {B,R,O,a, \lambda / {\lambda_{ij}}, \bm{\theta} ,\varepsilon } \right.)\\
 = \frac{{P(R,O,a,B,\lambda, \bm{\theta} ,\varepsilon )}}{{P(R,O,a,B,\lambda / \lambda_{ij}, \bm{\theta} ,\varepsilon )}}
\end{array}
\label{n_ij_condition}
\end{equation}

The numerator in Eq.~\ref{n_ij_condition} can be unfolded as:
\begin{eqnarray}
&&P(R,O,a,B,\lambda,\bm{\theta} ,\varepsilon )\nonumber\\
 &=& {\prod\nolimits_{n = 1}^N {\prod\nolimits_{m = 1}^M {P({r_{ij}}\left| {{\o _{ij}},{a_i},\theta ,\varepsilon } \right.)P({\o _{ij}}\left| {{B_j},{\lambda _{ij}}} \right.)} } }
 \label{n_num}
\end{eqnarray}
while the denominator in Eq.~\ref{n_ij_condition} can be unfolded as:
\begin{eqnarray}
&&P(R,O,a,B,\lambda /{\lambda _{ij}}, \bm{\theta} ,\varepsilon )\nonumber\\
& =& \left[ {\prod\nolimits_{n = 1}^{N/j} {\prod\nolimits_{m = 1}^M {P({R_{mn}}\left| {{O_{mn}},{a_m},\bm{\theta} ,\varepsilon } \right.)P({O_{mn}}\left| {{B_n},{\lambda _{mn}}} \right.)} } } \right] \times \nonumber\\
&&\left[ {\prod\nolimits_{m = 1}^{M/i} {P({R_{mj}}\left| {{O_{mj}},{a_m},\bm{\theta} ,\varepsilon } \right.)P({O_{mj}}\left| {{B_j},{\lambda _{mj}}} \right.)} } \right] \times \nonumber\\
&&\sum\nolimits_{{\lambda _{ij}}} {P({R_{ij}}\left| {{O_{ij}},{a_i},\bm{\theta} ,\varepsilon } \right.)P({O_{ij}}\left| {{B_j},{\lambda _{ij}}} \right.)} 
\label{n_den}
\end{eqnarray}
By substituting the numerator and denominator in Eq.~\ref{n_ij_condition} with Eq.~\ref{n_num} and~\ref{n_den} then eliminating the redundant terms, we get:
\begin{eqnarray}
&&P({\lambda _{ij}}\left| {B,R,O,a,\lambda /{\lambda _{ij}},\bm{\theta} ,\varepsilon } \right.)\nonumber\\
& = &\frac{{P({\o _{ij}}\left| {{B_j},{\lambda _{ij}}} \right.)}}{{\sum\nolimits_{{\lambda _{ij}}} {P({\o _{ij}}\left| {{B_j},{\lambda _{ij}}} \right.)} }}\nonumber\\
& =& \frac{{\displaystyle \frac{{\Gamma ({\alpha _{ij}} + {\beta _{ij}} + {\gamma _{ij}} + 1)}}{{\Gamma ({\alpha _{ij}} + 1)\Gamma ({\beta _{ij}} + 1)\Gamma ({\gamma _{ij}} + 1)}}b_j^{{\alpha _{ij}}}d_j^{{\beta _{ij}}}n_j^{{\gamma _{ij}}}}}{{\sum\nolimits_{{\lambda _{ij}}} {\left[ {\displaystyle \frac{{\Gamma ({\alpha _{ij}} + {\beta _{ij}} + {\gamma _{ij}} + 1)}}{{\Gamma ({\alpha _{ij}} + 1)\Gamma ({\beta _{ij}} + 1)\Gamma ({\gamma _{ij}} + 1)}}b_j^{{\alpha _{ij}}}d_j^{{\beta _{ij}}}n_j^{{\gamma _{ij}}}} \right]} }}
\label{n_final}
\end{eqnarray}
where $\lambda_{ij} = {{\alpha _{ij}} + {\beta _{ij}} + {\gamma _{ij}}}$ and $\sum\nolimits_{{\lambda _{ij}}}$ denotes the summation for all possible values of $\lambda_{ij}$.
Eq.~\ref{n_final} is solution for $P(\lambda\left| {R,O,a,B,\bm{\theta} ,\varepsilon } \right.)$.

\subsubsection{$P(a\left| {O,R,a,B,\lambda, \bm{\theta} ,\varepsilon } \right.)$}

%Eq.~\ref{n_ij_condition} can be further unfolded as:
%
%\begin{eqnarray}
%&&\frac{{P(R,O,a,B,n,\theta ,\varepsilon )}}{{P(R,O,a,B,n/{n_{ij}},\theta ,\varepsilon )}}\nonumber\\
 %&=& \frac{{\prod\limits_{n = 1}^N {\prod\limits_{m = 1}^M {P({R_{mn}}\left| {{O_{mn}},{a_m},\theta ,\varepsilon } \right.)P({O_{mn}}\left| {{B_n},{n_{mn}}} \right.)} } }}{{\prod\limits_{n = 1}^{N/j} {\prod\limits_{m = 1}^M {P({R_{mn}}\left| {{O_{mn}},{a_m},\theta ,\varepsilon } \right.)P({O_{mn}}\left| {{B_n},{n_{mn}}} \right.)} } \left( {\prod\limits_{m = 1}^{M/i} {P({R_{mj}}\left| {{O_{mj}},{a_m},\theta ,\varepsilon } \right.)P({O_{mj}}\left| {{B_j},{n_{mj}}} \right.)} } \right)\sum\limits_{{n_{ij}}} {P({R_{ij}}\left| {{O_{ij}},{a_i},\theta ,\varepsilon } \right.)P({O_{ij}}\left| {{B_j},{n_{ij}}} \right.)} }}\nonumber\\
%& = &\frac{{P({O_{ij}}\left| {{B_j},{n_{ij}}} \right.)}}{{\sum\limits_{{n_{ij}}} {P({O_{ij}}\left| {{B_j},{n_{ij}}} \right.)} }}\nonumber\\
% &= &\frac{{\frac{{\Gamma (n_{ij}^{} + 1)}}{{\prod\nolimits_{k = 1}^3 {\Gamma (n_{ij}^{(k)} + 1)} }}\prod\limits_{k = 1}^K {{{(B_j^{(k)})}^{n_{ij}^{(k)}}}} }}{{\sum\limits_{{n_{ij}}} {\left[ {\frac{{\Gamma (n_{ij}^{} + 1)}}{{\prod\nolimits_{k = 1}^3 {\Gamma (n_{ij}^{(k)} + 1)} }}\prod\limits_{k = 1}^K {{{(B_j^{(k)})}^{n_{ij}^{(k)}}}} } \right]} }}
 %\label{n_ij_final}
%\end{eqnarray}
%
%where $\sum\limits_{{n_{ij}}} {} $ denotes the summation for all possible values of $n_{ij}$.
%

For $P(a\left| {O,R,a,B,\lambda, \bm{\theta} ,\varepsilon } \right.)$, since $a$ is a vector where an entry is denoted as $a_{i}$, we need to further solve out:
\begin{eqnarray}
P({a_i}\left| {B,R,O,a/{a_i},\lambda, \bm{\theta} ,\varepsilon } \right.)
\label{a_i}
\end{eqnarray}
for each $a_{i} \in a$, where ${a/{a_{i}}}$ denotes all the other entries in $a$ except $a_{i}$.
According to Bayes' rule, Eq.~\ref{a_i} can also be expressed as:
\begin{equation}
\begin{array}{l}
P({a_i}\left| {B,R,O,a/{a_i},n,\theta ,\varepsilon } \right.)\\
 = \frac{{P(R,O,a,B,\lambda ,\theta ,\varepsilon )}}{{P(R,O,a/{a_i},B,\lambda ,\theta ,\varepsilon )}}
\end{array}
\label{a_i_condition}
\end{equation}

The numerator in Eq.~\ref{a_i_condition} can be unfolded as:
\begin{eqnarray}
&&P(R,O,a,B,\lambda,\bm{\theta} ,\varepsilon )\nonumber\\
 &=& {\prod\nolimits_{n = 1}^N {\prod\nolimits_{m = 1}^M {P({r_{ij}}\left| {{\o _{ij}},{a_i},\theta ,\varepsilon } \right.)P({\o _{ij}}\left| {{B_j},{\lambda _{ij}}} \right.)} } }
 \label{a_num}
\end{eqnarray}
while the denominator in Eq.~\ref{n_ij_condition} can be unfolded as:
\begin{eqnarray}
&&\frac{{P(R,O,a,B,\lambda ,\theta ,\varepsilon )}}{{P(R,O,a/{a_i},B,\lambda ,\theta ,\varepsilon )}}\nonumber\\
 &=& \left[ {\prod\nolimits_{m = 1}^{M/i} {\prod\nolimits_{n = 1}^N {P({R_{mn}}\left| {{O_{mn}},{a_m},\theta ,\varepsilon } \right.)P({O_{mn}}\left| {{B_n},{\lambda _{mn}}} \right.)} } } \right] \times \nonumber\\
&&\left[ {\int {\prod\nolimits_{n = 1}^N {P({R_{in}}\left| {{O_{in}},{a_i},\theta ,\varepsilon } \right.)P({O_{in}}\left| {{B_n},{\lambda _{in}}} \right.)d{a_i}} } } \right]\\
&=& \left[ {\prod\nolimits_{m = 1}^{M/i} {\prod\nolimits_{n = 1}^N {P({R_{mn}}\left| {{O_{mn}},{a_m},\theta ,\varepsilon } \right.)P({O_{mn}}\left| {{B_n},{\lambda _{mn}}} \right.)} } } \right] \times\nonumber \\
&&\left[ {\prod\nolimits_{n = 1}^N {P({O_{in}}\left| {{B_n},{\lambda _{in}}} \right.)} } \right] \times\nonumber \\
&&\int {\prod\nolimits_{n = 1}^N {P({R_{in}}\left| {{O_{in}},{a_i},\theta ,\varepsilon } \right.)d{a_i}} } 
\label{a_den}
\end{eqnarray}
By substituting the numerator and denominator in Eq.~\ref{a_i_condition} with Eq.~\ref{a_num} and~\ref{a_den} then eliminating the redundant terms, we get:
\begin{eqnarray}
&&P({a_i}\left| {B,R,O,a/{a_i},n,\theta ,\varepsilon } \right.)\nonumber\\
& =& \frac{{P(R,O,a,B,n,\theta ,\varepsilon )}}{{P(R,B,a/{a_i},B,n,\theta ,\varepsilon )}}\nonumber\\
 &=& \frac{{\displaystyle \prod\nolimits_{n = 1}^N {P({R_{in}}\left| {{O_{in}},{a_i},\theta ,\varepsilon } \right.)} }}{{\displaystyle \int {\prod\nolimits_{n = 1}^N {P({R_{in}}\left| {{O_{in}},{a_i},\theta ,\varepsilon } \right.)} d{a_i}} }}\nonumber\\
& = &\frac{{\displaystyle\prod\nolimits_{n = 1}^N {{\rm{logit}}\left[ {E({\omega _{in}}),\theta ,\varepsilon } \right]} }}{{\displaystyle\int {\prod\nolimits_{n = 1}^N {{\rm{logit}}\left[ {E({\omega _{in}}),\theta ,\varepsilon } \right]d{a_i}} } }}
\label{a_i_final} 
\end{eqnarray}
which is the solution for $P(a_i \left| {O,R,a/a_i,B,\lambda, \bm{\theta} ,\varepsilon } \right.)$.
Since the integration term in Eq.~\ref{a_i_final} does not have an analytic form, we use a numerical result to represent it.  

\subsubsection{$P(\varepsilon \left| {B,R,O,a,\lambda,\bm{\theta}} \right.)$}
For $P(\varepsilon \left| {B,R,O,a,\lambda,\bm{\theta}} \right.)$, since $\varepsilon$ is a single value, we directly apply Bayes' rule on it:
\begin{eqnarray}
&&P(\varepsilon \left| {B,R,O,a,\lambda ,\theta ,} \right.)\nonumber\\
 &=& \frac{{P(R,O,a,B,\lambda ,\theta ,\varepsilon )}}{{P(R,O,a,B,\lambda ,\bm{\theta} )}}
 \label{eps_condition}
\end{eqnarray}

The numerator in Eq.~\ref{eps_condition} can be unfolded as:
\begin{eqnarray}
&&P(R,O,a,B,\lambda,\bm{\theta} ,\varepsilon )\nonumber\\
&=& {\prod\nolimits_{n = 1}^N {\prod\nolimits_{m = 1}^M {P({r_{ij}}\left| {{\o _{ij}},{a_i},\bm{\theta} ,\varepsilon } \right.)P({\o _{ij}}\left| {{B_j},{\lambda _{ij}}} \right.)} } }
\label{eps_num}
\end{eqnarray}
while the denominator in Eq.~\ref{eps_condition} can be unfolded as:
\begin{eqnarray}
&&P(R,O,a,B,\lambda ,\theta )\nonumber\\
& =& \int {\prod\nolimits_{n = 1}^N {\prod\nolimits_{m = 1}^M {P({r_{ij}}\left| {{\o _{ij}},{a_i},\theta ,\varepsilon } \right.)P({\o _{ij}}\left| {{B_j},{\lambda _{ij}}} \right.)} } d\varepsilon } \nonumber\\
 &=& \left[ {\prod\nolimits_{n = 1}^N {\prod\nolimits_{m = 1}^M {P({\o _{ij}}\left| {{B_j},{\lambda _{ij}}} \right.)} } } \right] \times \nonumber\\
&&\int {\prod\nolimits_{n = 1}^N {\prod\nolimits_{m = 1}^M {P({r_{ij}}\left| {{\o _{ij}},{a_i},\bm{\theta} ,\varepsilon } \right.)} } d\varepsilon } 
\label{eps_den}
\end{eqnarray}
By substituting the numerator and denominator in Eq.~\ref{eps_condition} with Eq.~\ref{eps_num} and~\ref{eps_den} then eliminating the redundant terms, we get:
\begin{eqnarray}
&&P(\varepsilon \left| {B,R,O,a,\lambda ,\bm{\theta} } \right.)\nonumber\\
 &=& \displaystyle \frac{{P(R,O,a,B,\lambda ,\bm{\theta} ,\varepsilon )}}{{P(R,O,a,B,\lambda ,\bm{\theta} )}}\nonumber\\
 &=& \frac{{\displaystyle \prod\nolimits_{n = 1}^N {\prod\nolimits_{m = 1}^M {P({r_{ij}}\left| {{\o _{ij}},{a_i},\bm{\theta} ,\varepsilon } \right.)} } }}{{\displaystyle \int {\prod\nolimits_{n = 1}^N {\prod\nolimits_{m = 1}^M {P({r_{ij}}\left| {{\o _{ij}},{a_i},\bm{\theta} ,\varepsilon } \right.)} } d\varepsilon } }}
\label{eps_final} 
\end{eqnarray}
Eq.~\ref{eps_final} is the solution to  $P(\varepsilon \left| {B,R,O,a,\lambda,\bm{\theta}} \right.)$.
%\begin{eqnarray}
%&&P(\varepsilon \left| {B,R,O,a,n,\theta } \right.)\nonumber\\
 %&=& \frac{{P(R,O,a,B,n,\theta ,\varepsilon )}}{{P(R,B,a,B,n,\theta )}}\nonumber\\
 %&=& \frac{{\prod\limits_{m = 1}^M {\prod\limits_{n = 1}^N {P({R_{ij}}\left| {{O_{ij}},{a_i},\theta ,\varepsilon } \right.)P({O_{ij}}\left| {{B_j},{n_{ij}}} \right.)} } }}{{ \displaystyle \int {\prod\limits_{m = 1}^M {\prod\limits_{n = 1}^N {P({R_{ij}}\left| {{O_{ij}},{a_i},\theta ,\varepsilon } \right.)P({O_{ij}}\left| {{B_j},{n_{ij}}} \right.)} } d\varepsilon } }}\nonumber\\
 %&=& \frac{{\left( {\prod\limits_{m = 1}^M {\prod\limits_{n = 1}^N {P({O_{ij}}\left| {{B_j},{n_{ij}}} \right.)} } } \right)\prod\limits_{m = 1}^M {\prod\limits_{n = 1}^N {P({R_{ij}}\left| {{O_{ij}},{a_i},\theta ,\varepsilon } \right.)} } }}{{\left( {\prod\limits_{m = 1}^M {\prod\limits_{n = 1}^N {P({O_{ij}}\left| {{B_j},{n_{ij}}} \right.)} } } \right) \displaystyle \int {\prod\limits_{m = 1}^M {\prod\limits_{n = 1}^N {P({R_{ij}}\left| {{O_{ij}},{a_i},\theta ,\varepsilon } \right.)} } d\varepsilon } }}\nonumber\\
 %&=& \frac{{\prod\limits_{m = 1}^M {\prod\limits_{n = 1}^N {{\rm{logit}}(O_{mn}^{(0)} + {a_m}O_{mn}^{(2)},\theta ,\varepsilon )} } }}{{\displaystyle \int {\prod\limits_{m = 1}^M {\prod\limits_{n = 1}^N {{\rm{logit}}(O_{mn}^{(0)} + {a_m}O_{mn}^{(2)},\theta ,\varepsilon )} } d\varepsilon } }}
 %\label{epsilon_final}
%\end{eqnarray}
%
Similar as Eq.~\ref{a_i_final}, Eq.~\ref{epsilon_final} does not have a analytic form, we use a numerical solution to represent it.
\subsubsection{$ P(\bm{\theta} \left| {B,R,O,a,\lambda,\varepsilon} \right.)$}
For $P(\bm{\theta} \left| {B,R,O,a,\lambda,\varepsilon } \right.)$, Since $\bm{\theta}$ is a vector where an entry is denoted as $\theta_l (1 \le l < L)$ for $L$ possible levels, we need to further solve:
\begin{eqnarray}
P({\theta _l}\left| {B,R,O,a,\lambda ,\varepsilon } \right.)
\end{eqnarray}
for each $l$.
Similar as $\varepsilon$, $\bm{\theta}$ are also parameters of ordered logit model.
Therefore, $P({\theta _l}\left| {B,R,O,a,\lambda ,\varepsilon } \right.)$ has a similar form as Eq.~\ref{eps_final}, which can be expressed as:
\begin{eqnarray}
&&P({\theta _l}\left| {B,R,O,a,\lambda ,\bm{\theta} /{\theta _l}} \right.,\varepsilon )\nonumber\\
& =& \frac{{P(R,O,a,B,\lambda ,\bm{\theta} ,\varepsilon )}}{{P(R,O,a,B,\lambda ,\bm{\theta} /{\theta _l},\varepsilon )}}\nonumber\\
& =& \frac{{\displaystyle \prod\nolimits_{n = 1}^N {\prod\nolimits_{m = 1}^M {P({r_{ij}}\left| {{\o _{ij}},{a_i}, \theta_l ,\varepsilon } \right.)} } }}{{\displaystyle \int {\prod\nolimits_{n = 1}^N {\prod\nolimits_{m = 1}^M {P({r_{ij}}\left| {{\o _{ij}},{a_i}, \theta_l ,\varepsilon } \right.)} } d{\theta _l}} }}\nonumber\\
&=& \frac{{\displaystyle \prod\nolimits_{n = 1}^N {\prod\nolimits_{m = 1}^M {{\rm{logit}}\left[ {\varepsilon E({\omega _{ij}}) + {\theta _l}} \right]} } }}{{\displaystyle \int {\prod\nolimits_{n = 1}^N {\prod\nolimits_{m = 1}^M {{\rm{logit}}\left[ {\varepsilon E({\omega _{ij}}) + {\theta _l}} \right]} } d{\theta _l}} }}
\label{theta_final}
\end{eqnarray}

%Since $\bm{\theta}$ is a vector where an entry is denoted as $\theta_l (1 \le l < L)$ for $L$ possible levels, we need to further solve:
%
%\begin{eqnarray}
%P({\o_{ij}}\left| {R,O/{\o_{ij}},a,B,\lambda,\bm{\theta} ,\varepsilon } \right.)
%\label{O_ij}
%\end{eqnarray}
%   
%for each $\o_{ij} \in O$, where ${O/{\o_{ij}}}$ denotes all the other entries in $O$ except $\o_{ij}$.

%\begin{eqnarray}
%&&P(\theta \left| {B,R,O,a,n,\varepsilon } \right.)\nonumber\\
 %&=& \frac{{\prod\limits_{m = 1}^M {\prod\limits_{n = 1}^N {{\rm{logit}}(O_{mn}^{(0)} + {a_m}O_{mn}^{(2)},\theta ,\varepsilon )} } }}{{\displaystyle \int {\prod\limits_{m = 1}^M {\prod\limits_{n = 1}^N {{\rm{logit}}(O_{mn}^{(0)} + {a_m}O_{mn}^{(2)},\theta ,\varepsilon )} } d\theta } }}
%\label{theta_final}
%\end{eqnarray} 
\subsubsection{Gibbs Sampling Algorithm}

Combining Eq.~\ref{B_final},~\ref{O_ij_final},~\ref{n_final},~\ref{a_i_final}, ~\ref{eps_final} and~\ref{theta_final}, the Gibbs sampling algorithm to infer our PGM can be obtained, as shown in Algorithm~\ref{A2}

\begin{algorithm}
\label{A2}
\caption{Gibbs Sampling Algorithm for OpinionEst}
\begin{algorithmic}[1]
\REQUIRE Randomly initialized variables ${B^{(0)}},{O^{(0)}},{a^{(0)}},{\lambda^{(0)}},{\bm{\theta} ^{(0)}},{\varepsilon ^{(0)}}$.
\ENSURE The distribution of data samples $\left\{ {{B^{(t)}},{O^{(t)}},{a^{(t)}},{\lambda^{(t)}},{\bm{\theta} ^{(t)}},{\varepsilon ^{(t)}}} \right\}$ converge when $t \to \infty$
\FORALL  {iteration number $t = 1,2,3...$}
\FORALL  {$B_j \in B$}
\STATE Sample $B_j$ from Eq.~\ref{B_final} with parameters ${R^{}},{O^{(t - 1)}},{a^{(t - 1)}},{\lambda^{(t - 1)}},{\bm{\theta}^{(t - 1)}},{\varepsilon ^{(t - 1)}}$
\ENDFOR
\FORALL  {$\o_{ij} \in O$}
\STATE Sample $\o_{ij}$ from Eq.~\ref{O_ij_final} with parameters ${R^{}}, {B^{(t)}}, {a^{(t - 1)}},{n^{(t - 1)}},{\bm{\theta}^{(t - 1)}},{\varepsilon ^{(t - 1)}}$
\ENDFOR
\FORALL  {$a_i \in a$}
\STATE Sample $a_i$ from Eq.~\ref{a_i_final} with parameters $R,{B^{(t)}},{O^{(t)}},{\lambda^{(t - 1)}},{\bm{\theta}^{(t - 1)}},{\varepsilon ^{(t - 1)}}$
\ENDFOR
\FORALL  {$\lambda_{ij} \in \lambda$}
\STATE Sample $\lambda_{ij}$ from Eq.~\ref{n_final} with parameters $R,{B^{(t)}},{O^{(t)}},{a^{(t)}},{\bm{\theta}^{(t - 1)}},{\varepsilon ^{(t - 1)}}$
\ENDFOR
\FORALL  {$\theta_{l} \in \bm{\theta}$}
\STATE Sample $\theta_{l}$ from Eq.~\ref{theta_final} with parameters $R,{B^{(t)}},{O^{(t)}},{a^{(t)}},{\lambda^{(t)}},{\varepsilon ^{(t - 1)}}$, $\bm{\theta}^{(t-1)}/\theta_{l}^{(t-1)}$
\ENDFOR
\STATE Sample $\varepsilon$ from Eq.~\ref{eps_final} with parameters $R,{B^{(t)}},{O^{(t)}},{a^{(t)}},{\lambda^{(t)}},{\bm{\theta}^{(t)}}$
\ENDFOR
\end{algorithmic}
\label{GS_arch}
\end{algorithm}